\documentclass[conference]{IEEEtran}
\ifCLASSINFOpdf
\else
\fi
\usepackage{graphicx}
\usepackage{caption}

\usepackage{graphicx}
\usepackage[export]{adjustbox}

\usepackage[french,english]{babel}
\usepackage[T1]{fontenc}
\usepackage[utf8]{inputenc}
\usepackage{amsmath, amssymb}
\usepackage{comment}

\begin{document}
%
\title{Fusion of ANN and SVM Classifiers for Network Attack Detection}

\author{\IEEEauthorblockN{Takwa Omrani}
\IEEEauthorblockA{National School of \\ Engineering. \\ Université de 
Gabès. \\Tunisia.\\
 omranitakwa@yahoo.fr}
\and
\IEEEauthorblockN{ Adel Dallali}
\IEEEauthorblockA{Faculté de sciences de Gafsa.\\ Université de Gafsa. \\ Tunisia. \\
 adel.dallali@gmail.com}
\and
\IEEEauthorblockN{Bilgacem Chibani Rhaimi}                        
\IEEEauthorblockA{National School of Engineering\\Université de Gabès.\\
Tunisia.\\
 abouahmed97@gmail.com}
\and
\IEEEauthorblockN{Jaouhar Fattahi }
\IEEEauthorblockA{Department of Computer \\Science \& Software Engineering,\\ Université Laval, Québec, Canada\\ jaouhar.fattahi.1@ulaval.ca}}


%
\maketitle

\begin{abstract}
With the progressive increase of network application and electronic devices (computers, mobile phones, android, etc.) attack and intrusion,  detection has become a very challenging task in cybercrime detection area. in this context, most of the existing approaches of attack detection rely mainly on a finite set of attacks. These solutions are vulnerable, that is, they fail in detecting  some attacks when sources of informations are ambiguous or imperfect. However, few approaches started investigating in this direction. This paper investigates the role of machine learning approach (ANN, SVM) in detecting a TCP connection traffic as a normal or a suspicious one. But, using ANN and SVM  is an expensive technique individually. In this paper, combining two classifiers are proposed, where artificial neural network (ANN) classifier and support vector machine (SVM) are both employed. Additionally, our proposed solution allows to visualize obtained classification results. Accuracy of the proposed solution has been compared with other classifier results. Experiments have been conducted with different network connections selected from NSL-KDD DARPA dataset. Empirical results show that combining  ANN and SVM techniques  for attack detection is a promising direction.\\
\end{abstract}
\begin{IEEEkeywords}
Attack, Detection, Classification, ANN, SVM, Fusion.
\end{IEEEkeywords}


%
\IEEEpeerreviewmaketitle

\section{Introduction}
 With the phenomenal growth of the Internet connection, many forms of cybercrimes are growing continuously.As example, theft of data, fishing, carding, viruses, financial fraud, intrusions and attacks are potential forms of cybercrimes \cite{3} which create from its exploration a challenging task. Network attacks is  one of cybercrime types which intends to compromise the confidentiality, integrity and/or availability of the information be it  in  the network traffic or in the local host \cite{1}.\\

 With the continuous number and forms of attacks, there has been a huge number of Internet Detection Systems (IDSs) identifying malicious attacks in order to protect computer systems from possible damages. However,  some  unknown attacks remained hard challenges\cite{2} for IDS. Sometimes, where the source of data is ambiguous or imperfect, IDS cannot able to differentiating between normal and suspicious connection \cite{5}. For that, attack detection accuracy of system can be degraded and  the false positive rate of detection stays increased  even though accuracy stays low\cite{4}.\\

To improve attack detection rate and to facilitate
administration of IDS, many approaches have been
suggested in the literature that can offers a major
opportunity \cite{4,5} in detection and classification
network connection for unknown attacks. One
approach is machine learning (ML) which builds
models from training data because anomaly
and intrusion detection can be treated as a
classification challenge.\\

Even though the classification features are
ambiguous or imperfect, some classifiers, using
random factors, can generate higher overall
accuracy of detection. Therefore, it is necessary to move from a certain environment to an uncertain attack detection environment. This idea has been the basic objective of our classic data fusion system. This helps to combine information to improve the decision making of an information with a reduced error rate. \\

Following this trend, some works emerged and
explored a new attack detection approach with
using an hybrid approach that merges classification algorithms. But, these works still suffer from the lack of dealing with ambiguous sources of
information.\\

To resolve this problem, we propose, in this
paper to improve the complexity of supervised
and unsupervised ML techniques in detecting
attack and intrusion process. In fact, a comparison
of SVM and ANN classifiers for network
intrusion detection will be presented and their
computational complexity will be discussed. Then,
to reduce error rate experimented by ANN and
SVM classifiers, we suggest applying classic data
fusion approach based conditional probability that
combine ANN and SVM classifier decision.
To validate our experimentation results, we
evaluated our approach on the real benchmark
of TCP connections gathered in DARPA KDD99
dataset \cite{15}. Our results show that combining
classifiers decisions and features selection can
reduce the error rate of attack detection.
The rest of the paper is organized as follows;
section II states different attack classification and
detection approaches applying ML techniques.
In section III, we show classification results of
ANN, SVM then we compare their performance.
Finally, we give a short conclusion and futures
works. 
 

 \section {Related Work}
Many approaches have been proposed to solve attack detection problem within network connection ranging from supervised approaches \cite{7} unsupervised approaches \cite{24} to hybrid ones \cite{25, 26}.\\

Supervised approach use a wide range of features and labeled data for training attack classifiers. For instance, Haddadi et al.\cite{6} proposed IDS using feed-forward neural network with back propagation algorithm for network based intrusion detection. This scheme deals with KDD-CUP’99 dataset for the classification of network attacks. In the same context, Haidar et al. \cite{14} introduced anomaly-based detection system  using supervised neural networks classifier to show the reliability of intrusion system.\\

The first issue is the luck of a training data set that increase the complexity of classification algorithm. Moreover, having significants features is a challenge and, where the information sources are unbalanced, the training sets contain some noises that result increase false alarm rates. \\

To overcome this problem, authors proposed unsupervised approaches to attack detection of network connection. For instance, Chandola et al. \cite{21} suggest a support vector machine (SVM) classifier for attack classification. In this context, Moor Andrew \cite{23} used the Markov model-based intrusion detection system to calculate the probability  of  attacks presented in the system based on a list of observations. For example, a sequence of alerts from an intrusion detection system (IDS) such as Snort can be used to calculate the probability of system being  attacked. Furthermore, Bronstein Alexandre et  al. \cite{22} propose networks model for network intrusion detection based on a Bayesian probability . In the same context of unsupervised method, Imam Riadi et al. \cite{9} used  K-means clustering technique was tested in order to range attack in TCP and UDP connection into three classes respectively known as very dangerous, rather dangerous and not dangerous attack but this method cannot deal with a huge number of intrusion scenarios.\\

However, unsupervised classifiers is based only on training of normal attack. Then, any deviation is judged as an attack. This algorithm accurate weakly in attack detection, that is, the number of false positive rate remains important.\\

But, using one classifier algorithm (supervised or unsupervised) to classify the network traffic data as normal behavior or anomalous, cannot give the best possible attack detection accuracy and cannot reduced alarm rate. \\

Hence, some hybrid approaches, that merge both supervised and unsupervised methods, are combined. For instance,  M.Elbasiony et al. \cite{25} proposed an hybrid IDS based on two famous data mining algorithms called random forests and k-means. In the same context, Panda et al. \cite{ 26} proposed an hybrid attack detection algorithm that  combine Decision Trees, SVM  and Random Forest to  classify network attack.\\

However, these techniques suffer from identifying all intrusion attempts, that as, it cannot achieve a higher detection rate and lower false alarm rate. Moreover, Tsujii \cite{11} used the Naive Bayes classifier combined with the well-established EM algorithm to exploit the unlabeled data. In other works, Naïve Bayes and decision tree algorithm were combined for network intrusion detection which provided high accuracy for different types of network \cite{12}. \\

But all the works presented above use decisions of independent classifiers without emphasizing the dependence of the combined classes. Indeed, under the assumption that there is a gap between classes, the attack detection may be easier to implement, faster to evaluate and reduce the amount of training data needed to estimate the attack.\\

Hence, including feature selection and classifier techniques can achieve a better performance. Following this trend, our work tends to be placed where, classifiers and three feature selection (protocol, service and flag)  are combined. In our work, support vector machine (SVM) and artificial neural network (ANN)  have a higher classification accuracy  in comparison to other classifier models but due to the higher training time for large data sets, the usage is limited. Hence many feature selection techniques are integrated with SVM and ANN classifiers to have an accurate result of attack detection.\\

\section{Attack Detection Analysis of Network Connection}
\subsection{Data Description}
 We randomly selected a part samples containing 10000 samples refers to NSL-KDD dataset. NSL-KDD dataset is a subset of the KDD'99 TCP connection corpus developed in MIT Lincoln which is publicly available on \cite{15}. NSL-KDD dataset has been offered to overcome the  redundancy of previous version of DARPA datasets ( KDD’99, KDD'2000) that can reduce the accuracy of our experimentation \cite{14}. The labeled dataset is selected for  attack classification task. Presented attack  are grouped into four classes : (i) Denial of service (Dos), where some resource is flooded, causing DoS to safe users.(ii) Probes which is collecting network information to avoid security tools. (iii) Remote to Local (R2L) attacks that use remote system vulnerabilities to penetrate a system. (iiii) User to root (U2R) attacks that aims to gain root access to a system. \\

In our paper, we are interested at binary  classification of connection,  then two  labeled classes are presented; "attack" with score 1 and "normal" with score 0.\\

The selected subset consists of   125373 samples collected from the 58630 for attack class and 67343 for normal class \cite {15}.\\

\subsection{Data Preprocessing }
Before starting the classification process, a preliminary cleaning step of the database used is required. This stage consists of four main phases\\
\begin{enumerate}
\item  More than 50\% 
of all instances are attacks. thus, in order to achieve a reduced attack rate and a balanced distribution of normal and attack classes. we must adjust the number of attacks; 50\% of the connecion is selected as the normal connection and 50\% as the attack connection;

\item We can notice that the distribution of the connection states in the base NSL-KDD is not balanced. It is dominated formally by the probes (11656 instances). Denial of Service (DOS) attacks (45,927 instances) are also numerous and have millions of instances. For this it is necessary to remove some of these attacks to achieve balance with other types of attacks like R2L (52 instances) only and U2R (995);

\item The attack data is split into two disjoint partitions containing only attack and normal types. Therefore, we decomposed the selected data into three disjoint sets of equal length : the training data, the validation data and test data;

\item To perfectly express our proposed algorithm, we randomly selected 1000 instances for validation and testing. Then we imposed a 50\% attack rate to preserve the balanced distributions of attack types. An example of reducing and standardized data is shown in the table \ ref {features}.
\end{enumerate}

 \begin{table}
\centering
\begin{tabular}{|c|r|r|r|}
\hline
feature 5&feature 6&feature 32&class\\
\hline
 0.282 & 0.0351 & 0.9970 & 0\\
  0.3180 &	0.3305	& 1.3382 &1\\
\hline
0.0280 &	0.0462 & 1.2255 & 0\\
\hline
0.3180 & 0.3305	& 0.8434 &1\\
\hline
0.0280 & 0.1127	& 0.3632 &0\\
\hline
 0.0282 &	0.0714 & 1.3405& 0\\
\hline
1.4425 & 3.2210	& 1.4527 &1\\
\hline
\end{tabular}
\caption{ \label{features} example of used dataset }
\end{table}

\subsection{Features Extraction}

In our work, we need to two forms of features (low level feature and high level features). Low level feature is extracted directly from the dataset which are (IP destination, IP source, port number source and port number destination). Hight level features are calculated from  ANN and SVM classifiers decision.\\

Classifiers decision are presented in term of percentage of correct classification  (PCC). This value is calculated from matrix covariance of compared classifiers as presented in table \ref{matrix}. PCC is presented by formula (1).\\

\[ PCC=\frac{TP+TN}{TP+TN+FP+FN} \]

\begin{table}
\centering
\begin{tabular}{|c|r|r|}
\hline
actual class & normal class &  attack class\\
\hline
normal class & TP &FP\\
\hline
attack class & FN & TN\\
\hline
\end{tabular}
\caption{ \label{matrix} Covariance Matrix}
\end{table}
where:\\
•	TP – Number of samples which is well classified as normal.\\
•	TN – Number of samples which is well classified as Intrusion.\\
•	FP – Number of samples classified as Intrusion but they were normals.\\
•	FN – Number of samples classified as Normal but they were attacks.\\
To evaluate the performance of classifiers, two criteria were used as demonstrated namely TP rates and FP rates.\\

\section{Binary Classification of Network Connections}

The fundamental task in this paper is for estimation of network connection as normal or positive.
Therefore, for this task we proposed a binary classier using Naive Bayes (NB) classifier. \\

To experiment this BN classier, we use the following formal model. Let ${f_{1},...,f_{k}}$ be a predefined set of k features (PCC) presented  in network connection. Each feature $f_{i}$ could be expressed in term of percentage of correct classification (PCC) given by each individual classifier.\\

Let $w_{i}(r)$ be the function that showing how the feature $f_{i}$ occurs in the network connection $nc$. Then, each  instance of connection  is listed by the following review vector:\\

\[ W_{nc}=(w_{1}(nc);w_{2}(nc),...,w_{k}(nc))\]
In the proposed  network attack detection process,we  assign to a each network connection $nc$ the class $cl$.
We conduct our NB classier by first observing that by Bayes'rule:\\

\[  P(cl\backslash nc)=\frac{P(cl)P(nc\backslash cl)}{P(nc)}\]

Where $P(nc)$ hasn't any role  in selecting $cl$. The term $P(nc\backslash cl)$ is estimating by Naive Bayes that decomposed it by assuming the $fi$'s are conditionally independent given $n$'s class:\\
\[ P_{NB}(cl\backslash nc)=\frac{P(cl)\prod_{i=1}^{k}P(f_{i}\backslash cl)^{n_{i}(nc)}}{P(nc)}\]

We implement our network attack classier in JAVA using an open source code.

\section{Evaluation Results}
In this section, we evaluate the fusion of both ANN and SVM classifiers decisions on binary attack 
detection and present the obtained results 	(PCC : percentage of correct class) on our network connection dataset. We then compare these results with those obtained using one decision class.\\

The objective of our paper is the evaluation effectiveness of the multiple decision classifiers for  network attack analysis. How useful are the data fusion features on network attacks connections ? How much gain do we get from combining classifiers decisions?\\

For our set of experiments, we use a probabilistic model based on NB method. Our model is  trained from preprocessed NSL-KDD samples as a baseline source. In each experimentation, we incorporate decision classes features into NB by either modifying the classifier decisions (PCC) techniques with merging other selected features that characterize the packet TCP of network connection, which can be flag, service or/and protocol. \\

Table \ref{NB} shows given classification results of our  binary attack detection  using  ML classifiers (ANN, SVM, and the combinations of decision classifiers through variation and replacement of three selected features into Naive Bayes).\\
\begin{table}
\begin{tabular}{|c|r|r|}
\hline
classifier&TPR&FPR\\
\hline
ANN&79.56&1.2 \\
\hline
SVM&79.27&1.4\\
\hline
ANN+SVM &79.65&1.1\\
\hline
ANN+SVM+ (flag and protocol features)&\textbf{79.71}&0.92\\
\hline
ANN+SVM+(service and protocol feature)&79.63&\textbf{0.75}\\
\hline
ANN+SVM+(flag, service and protocol feature)&79.58&1.0\\
\hline
\end{tabular}
\caption{ \label{NB} results of NSL-KDD dataset using several combinations of classifier's decisions.}
\end{table}
According to the results shown in Table \ref{NB}, the incorporation of fusion of  classifier's decisions based on NB method outperforms the classification model in term of increasing detection rate  and reducing error rate.\\

The gain is small and that is something we were expecting as NSL-KDD dataset is of a specific domain of attack detection and combined classifiers are not able to well detect each network connection attacks.\\

\section{Conclusion}
In this paper, we presented a data fusion approach for attack detection in network connection dataset using the conditional probability represented by $p$ naive Bayes method. We defined features of classification in term of PCC of classifier $cl$  calculated with ANN and SVM classifiers.\\

Then, we trained an attack detector that is able to determine positive and negative
network connection. The classier is based on the multinomial Naive Bayes classier that uses  ANN and SVM classifiers decisions combining with other selected features from NSL-KDD dataset. Hence, based on PCC of classifiers (ANN and SVM) and  three selected features (flag, protocol and service), we create a novel probabilistic fusion model that combine both of high and low level features.\\

As future work, we want to model the problem of attack analysis in network connections as multi-class classification problem. In fact, it is possible to classify the attack  in
more than two classes like "DOS", "BROB", "R2L", etc.\\

Furthermore, we plan to consider this problem as a regression problem as we can estimate the
degree of affinity for the attack instead of a simple negative/positive class.\\

As well as, we plan to perform more experiments on different type of random user-generated data other than NSL-KDD.\\

As far as we know, the problem of classifying attacks based on different sources of information cannot be solved in purely supervised and unsupervised techniques. A data fusion technique offers a promising direction for future research.\\




%
\bibliographystyle{IEEEtran}
\bibliography{referencesp1}

\section*{Notice}
© 2018 IEEE. Personal use of this material is permitted. Permission from IEEE must be obtained for all other uses, in any current or future media, including reprinting/republishing this material for advertising or promotional purposes, creating new collective works, for resale or redistribution to servers or lists, or reuse of any copyrighted component of this work in other works.

\end{document}